\documentclass{elsart}

\usepackage{epsfig}

\begin{document}

\begin{frontmatter}

\title{Polydisperse fluid mixtures of adhesive colloidal particles}

\author{Domenico Gazzillo\thanksref{mail}}
\author{and Achille Giacometti}

\address{INFM and Dipartimento di Chimica Fisica, \\
Universit\`{a} di Venezia, S. Marta DD 2137, I-30123 Venezia, Italy}

\thanks[mail]{Corresponding author. {\it E-mail address:} gazzillo@unive.it}

\date{\today}

\begin{abstract}
We investigate polydispersity effects on the average structure factor of
colloidal suspensions of neutral particles with surface adhesion. A sticky
hard sphere model alternative to Baxter's one is considered. The choice of
factorizable stickiness parameters in the potential allows a simple analytic
solution, within the ``mean spherical approximation'', for any number of
components and arbitrary stickiness distribution. Two particular cases are
discussed: i) all particles have different sizes but equal stickiness (Model
I), and ii) each particle has a stickiness proportional to its size (Model
II). The interplay between attraction and polydispersity yields a markedly
different behaviour for the two Models in regimes of strong coupling (i.e. 
strong adhesive forces and low temperature) and large polydispersity.
These results are then exploited to reanalyze experimental scattering
 data on sterically stabilized silica particles.


\end{abstract}

\end{frontmatter}

\section{Introduction}

In studies on colloidal suspensions of neutral particles with adhesive
interactions \cite{Robertus89,Duits90,Duits91}, several attempts have been
made to fit small angle scattering data by using Baxter's ``sticky hard
sphere'' (SHS) model \cite{Baxter68}. In addition to a hard sphere (HS)
repulsion, the SHS potential contains a surface adhesive term, introduced by
Baxter as a limit of an attractive square well which becomes infinitly deep
and narrow according to a particular prescription \cite{Baxter68}. For this
model the Ornstein-Zernike (OZ) integral equations of the liquid state
theory have been solved analytically within the Percus-Yevick (PY)
approximation, and expressions for the correlation functions are thus
available \cite{Baxter68,Perram75,Barboy79}. Unfortunately, however, for a
fluid with $p$ components the SHS-PY analytical solution requires the
knowledge of a set of density-dependent parameters $\lambda _{ij}$, to be
determined through $p(p+1)/2$ coupled quadratic equations \cite
{Perram75,Barboy79}. This feature makes Baxter's SHS-PY solution for
mixtures practically inapplicable to colloidal systems with significant
polydispersity, i.e. with large $p.$

Polydispersity means that the mesoscopic suspended particles of a same
chemical species may differ in size, charge or other properties.
Consequently, a polydisperse fluid is always a mixture with a large number
of components and strong asymmetries (in size, charge, etc.), even when all
meso-particles belong to a unique chemical species.

In a recent paper (hereafter referred to as Reference I) \cite{Gazzillo00}
we showed that the aforesaid shortcomings of the SHS-PY solution can be
avoided by using a different version of SHS potential, for which the OZ
equations are analytically solvable under a different closure, i.e. the
``mean spherical approximation'' (MSA) \cite{Brey87,Mier89,Tutschka98}. In
this alternative SHS model, the adhesive term is defined starting from a
Yukawa attractive tail and taking the ``sticky limit'' of infinite amplitude
and vanishing range, with their product remaining constant. In Ref. I it
was pointed out how, in the particular case of {\it factorizable stickiness
parameters} in the adhesive term, the SHS-MSA solution leads to closed
analytical expressions for scattering intensity and other ``global''
structure factors, which are valid for {\it arbitrary} $p$ and therefore are
suitable to fit experimental scattering data from polydisperse systems.

In the present work we shall review the main points of Ref. I, adding some
new results and remarks. We shall then apply the SHS-MSA solution to
small-angle-neutron-scattering data from a colloidal fluid formed by
sterically stabilized silica particles.

\section{Model and scattering functions}

Our SHS potential is defined by

\begin{equation}
\beta u_{ij}(r)=\beta u_{ij}^{{\rm HS}}(r)-K_{ij}\sigma _{ij}^{-1}\ \delta
_{+}(r-\sigma _{ij}),\qquad  \label{m1}
\end{equation}

\noindent with factorizable stickiness parameters \cite{Yasutomi96,Herrera98}

\begin{equation}
K_{ij}=Y_iY_j,\quad \quad Y_i\ge 0,\quad \quad (i,j=1,\ldots ,p),  \label{m2}
\end{equation}

\noindent which measure the strength of surface attraction. In Eq. (\ref{m1}%
), $u_{ij}^{{\rm HS}}(r)$ is the HS potential and the next term stems from
the ``sticky limit'' of a Yukawa tail \cite{Ginoza96}; $\beta =\left(
k_BT\right) ^{-1}$, $\sigma _{ij}=(\sigma _i+\sigma _j)/2,$ with $\sigma _i$
being the HS diameter of species $i$, and $\ \delta _{+}(x)$ is the
asymmetrical Dirac delta function defined by: $\int_a^bdx\ f(x)\delta
_{+}(x-c)=f\left( c^{+}\right) ,$ if $a\leq c<b$. The MSA solution for this
 model is represented by the factor correlation functions 

\begin{equation}
q_{ij}(r)=\left\{ 
\begin{array}{l}
\frac 12a_i(r^2-\sigma _{ij}^2)+b_i(r-\sigma _{ij})+K_{ij},\qquad L_{ij}\leq
r\leq \sigma _{ij} \\ 
0,\qquad \qquad \qquad \qquad {\rm elsewhere}
\end{array}
\right.  \label{m3}
\end{equation}
\noindent
where $L_{ij}=(\sigma _i-\sigma _j)/2$ and the expressions of $a_i$ and $b_i$
can be found in Ref. I. Once that the $q_{ij}(r)$ are known, all correlation
and scattering functions are in principle computable. The {\it partial }%
structure factors are given by \cite{Gazzillo00,Gazzillo97} 

\begin{equation}
S_{ij}(k)=\sum_{m=1}^p\widehat{Q}_{im}^{-1}\left( k\right) \widehat{Q}%
_{jm}^{-1}\left( -k\right) ,  \label{m4}
\end{equation}

\noindent where $\widehat{Q}_{ij}\left( k\right) =\delta _{ij}-2\pi (\rho
_i\rho _j)^{1/2}\widehat{q}_{ij}\left( k\right) ,$ with $\widehat{q}%
_{ij}\left( k\right) $ being the unidimensional Fourier transform of $%
q_{ij}(r)$ ($\delta _{ij}$ the Kronecker $\delta ,$ $\rho _i$ the number
density of species $i$). The scattering intensity as well as other {\it %
global} structure factors can then be calculated, all these scattering
functions being weighted sums of $S_{ij}(k),$ like $\sum_{i,j=1}^pw_i\left(
k\right) w_j^{*}\left( k\right) S_{ij}\left( k\right) $ (here, the asterisk means
complex conjugation). As an example, we mention the {\it measurable average\
structure factor} 
\begin{equation}
S_{{\rm M}}\left( k\right) =\sum_{i,j=1}^p\left( x_ix_j\right)
^{1/2}F_i(k)F_j^{*}(k)S_{ij}\left( k\right) \ /\sum_{i=1}^px_i\left|
F_i(k)\right| ^2,
\end{equation}
\noindent
where $x_i$ is the molar fraction of species $i$ and $F_i(k)$ its form
factor. Eq. (\ref{m4}) shows that the main difficulty in the computation of $%
S_{ij}(k)$ lies in the inversion of a $p\times p$ matrix $\widehat{{\bf Q}}%
(k),$ which usually becomes a formidable task with increasing $p.$
Nevertheless, as discussed in Ref. I, when $K_{ij}$ is factorizable as in
Eq. (\ref{m2}), $\widehat{{\bf Q}}(k)$ becomes a {\it dyadic} (or Jacobi)
matrix, i.e. $\widehat{Q}_{ij}(k)=\delta _{ij}+\sum_{\mu =1}^na_i^{(\mu
)}(k)b_j^{(\mu )}(k)$, which admits {\it analytic} inverse for {\it any }$p$
(in general, the same property does not hold for Baxter's SHS-PY solution 
\cite{Herrera91}). Another important consequence of the dyadic structure is
that it allows to compute the {\it global} scattering factors {\it bypassing}
a preliminary calculation of the individual $S_{ij}(k)$. Indeed, when $%
\widehat{{\bf Q}}(k)$ is dyadic the sums $\sum_{i,j=1}^pw_i\left( k\right)
w_j^{*}\left( k\right) S_{ij}\left( k\right) $ can be worked out
analytically (see Refs. \cite{Gazzillo00,Gazzillo97,Vrij79} for details),
unlike the most common case where they are performed numerically by
evaluating $p(p+1)/2$ independent contributions $S_{ij}(k)$ at each $k$.
When fitting experimental data, the availability of closed analytical
expressions directly for scattering intensity and $S_{{\rm M}}\left(
k\right) $ clearly represents a great advantage.

\section{Polydispersity effects in SHS fluids}

\subsection{Size and stickiness distributions}

To represent size polydispersity, we select $p$ possible diameters uniformly
distributed in an interval $(0,\sigma _{\max })$ with mesh size $\Delta
\sigma =0.02\left\langle \sigma \right\rangle $. Each diameter $\sigma _i$
characterizes a different component. Its molar fraction is $x_i=f(\sigma
_i)\Delta \sigma ,$ with $f$ expressed by a Schulz distribution, i.e. 
\begin{equation}
f(\sigma )=b^a\sigma ^{a-1}e^{-b\sigma }/\Gamma (a)\;\;\quad (a>1),
\label{p1}
\end{equation}
\noindent
where $\Gamma $ is the gamma function, $\left\langle \sigma \right\rangle $
the average diameter, $a=1/s_\sigma ^2$, $b=a/\left\langle \sigma
\right\rangle $ and $s_\sigma =[\left\langle \sigma ^2\right\rangle
-\left\langle \sigma \right\rangle ^2]^{1/2}/\left\langle \sigma
\right\rangle $ measures the degree of size polydispersity. We use $p=85$
components when $s_\sigma =0.1,$ and $p=175$ when $s_\sigma =0.3$ .

The model can in principle be worked out for any choice of $Y_i$, which meets
the requirements that $Y_i$ must be lenghts and $K_{ij}=Y_iY_j$ must be
proportional to $\beta =(k_BT)^{-1}$. However, two choices are particularly
interesting:

{\bf Model I }{\it (polydisperse in size but monodisperse in stickiness)},
which assumes that all particles have the same stickiness, i.e. 
\begin{equation}
Y_i=Y=\gamma _0\left\langle \sigma \right\rangle ,\qquad {\rm with}\qquad
\gamma _0^2=\frac{\varepsilon _0}{k_BT}=\frac 1{T^{*}}\ .  \label{p2}
\end{equation}
\noindent Here, $\gamma _0$ measures the strength of surface adhesion; $%
\varepsilon _0$ is an energy, and $T^{*}$ a reduced temperature. High $T^{*}$
(low $\gamma _0$) values correspond to weak surface adhesion or high
temperature; vice versa, low $T^{*}$ (high $\gamma _0$) values mean strong
attraction or low temperature.

{\bf Model II }{\it (with stickiness polydispersity linearly related to the
size one). }Since it is reasonable to expect that larger particles attract
each other more strongly, a simple way of introducing stickiness
polydispersity is the choice 
\begin{equation}
Y_i=\gamma _0\;\sigma _i\ .
\end{equation}

\subsection{Regimes for $S_{{\rm M}}(k)$ and generalized Boyle temperature}

Polydispersity effects on $S_{{\rm M}}(k)$ can be best described after
recalling some features of the monodisperse SHS structure factor.

In $S_{{\rm mono}}\left( k\right) $ the presence of strong adhesive forces
may be revealed mainly from its behaviour in the low-$k$ region. Here, $S_{%
{\rm mono}}^{{\rm SHS}}\left( k\right) $ may differ significantly from $S_{%
{\rm mono}}^{{\rm HS}}\left( k\right) ,$ exhibiting a drastic increase as $%
k\rightarrow 0$ (see Figure 1 for $s_\sigma =0$). More precisely, in Ref. I
we recognized the existence of two different ``regimes'' for $S_{{\rm mono}%
}^{{\rm SHS}}\left( k\right) $ respectively\ above and below the {\it Boyle
temperature,} $T_{{\rm B}}^{*}=3,$ where attractive and repulsive forces
balance each other in such a way that $B_2$ vanishes (the second virial
coefficient $B_2$ appears in the low-density expansion of $S_{{\rm mono}%
}\left( 0\right) =1-2B_2(T^{*})\rho +{\mathcal O}(\rho ^2)$ ). When $T^{*}>T_{%
{\rm B}}^{*}$ ($B_2>0$) the fluid behaves\ like pure HS without stickiness,
repulsive forces are dominant, and $S_{{\rm mono}}\left( 0\right) $ - which
is related to compressibility and density fluctuations - decreases with
increasing volume fraction $\eta $. \ When $T^{*}<T_{{\rm B}}^{*}$ ($B_2<0$) 
$S_{{\rm mono}}\left( 0\right) $ \ has a non-monotonic dependence on\ $\eta $%
. After an initial increase (which may become very strong, signalling\ the
approach to a gas-liquid phase transition with critical temperature $T_{{\rm %
c}}^{*}\simeq 1.61$), an inversion occurs and\ $S_{{\rm mono}}\left(
0\right) $ decreases with\ $\eta $. In other words, below $T_{{\rm B}}^{*}$
the attractive forces are dominant at low $\eta $, whereas repulsion
prevails at higher\ $\eta .$

For polydisperse SHS, there exist two similar ``regimes'' for $S_{{\rm M}%
}\left( k\right) $. The new feature is that a crossover between these
regimes can now be induced not only by a change of $T^{*}$, but also by a
change of polydispersity $s_\sigma $ (at fixed $T^{*}$).

Figure 1 illustrates the polydispersity effects on $S_{{\rm M}}(k),$
displaying its behaviour at $T^{*}=2.04$ ($\gamma _0=0.7$) and $\eta =0.2$,
as $s_\sigma $ increases from $0$ (monodisperse limit) to $0.1$ and $0.3.$
We employed form factors $F_i(k)$ for homogeneous spherical scattering cores
inside the particles, with scattering core diameters $\sigma _i^{{\rm scatt}%
} $ coincident with the HS ones, i.e. $\sigma _i^{{\rm scatt}}=\sigma _i$.
Polydispersity progressively dumps all oscillations in the first peak region
and beyond. Here, Model I and II are nearly equivalent, whereas their $S_{%
{\rm M}}(k)$ may strongly differ near the origin as a consequence of the
different stickiness distributions. In fact, increasing $s_\sigma $ (at the
fixed $T^{*}$ value) does not produce marked differences in Model II. On the
contrary, in Model I a large polydispersity can overwhelm the attractive
effects, leading to a strong HS-like decrease of $S_{{\rm M}}(0)$ and thus
preventing the possibility of an instability in the system.

In Ref. I this scenario was interpreted in terms of a {\it generalized Boyle
temperature, }$T_{{\rm B,F}}^{*}$. From the low-density expansion of $S_{%
{\rm M}}(0)=1-2B_{2,{\rm F}}(T^{*})\rho +{\mathcal O}(\rho ^2),$ we derived $%
B_{2,{\rm F}}$ as a generalization of $B_2$ when all form factors are
included. Then $B_{2,{\rm F}}(T_{{\rm B,F}}^{*})=0$ defines $T_{{\rm B,F}%
}^{*}(s_\sigma )$, which is a decreasing function of $s_\sigma $ (its
expression is given in Ref. I). Clearly, $T_{{\rm B,F}}^{*}(s_\sigma =0)=(T_{%
{\rm B}}^{*})_{{\rm mono}}=3.$ In a $(s_\sigma ,T^{*})$ diagram, above the
curve $T^{*}=T_{{\rm B,F}}^{*}(s_\sigma )$ there is the region corresponding
to a ``HS-like, repulsive regime''; below, there is the
``attractive-regime'' region. Now, as shown in the inset of Figure 1, $T_{%
{\rm B,F}}^{*}(s_\sigma )$ of Model II decreases very slowly, asymptotically
approaching $2.\,57;$ therefore, all three $(s_\sigma ,T^{*}=2.04)$-states
of Figure 1 lie in the lower region. On the other hand, $T_{{\rm B,F}%
}^{*}(s_\sigma )$ of Model I is a rapidly decreasing function, with $T_{{\rm %
B,F}}^{*}(s_\sigma =0.1)=2.79>T^{*}$ and $T_{{\rm B,F}}^{*}(s_\sigma
=0.3)=1.68<T^{*}$. This argument interprets the crossover from
``attractive'' to ``repulsive'' regime in terms of an increasing
polydispersity.

\section{Fit of experimental data}

To investigate polydispersity effects in real colloidal fluids, we now apply
our Models to experimental data for silica particles coated with octadecyl
chains and dispersed in benzene. For this system Duits {\it et al.} \cite
{Duits91} published a wide set of small-angle neutron-scattering results,
which exhibit a crossover from ``HS-like repulsive'' to ``attractive''
regime controlled by variation of $T$: on lowering the temperature,
attractive effects increase and a reversible phase separation into two
phases of different density occurs at $t=$ $32.5$ $^{\circ }$C. Duits {\it %
et al.} \cite{Duits91} fitted their experimental data using an
implementation by Robertus {\it et al.} \cite{Robertus89} of Baxter's SHS-PY
model, with stickiness independent of size and size polydispersity
represented by a rather limited number of components (up to $p=9$), due to
the aforesaid shortcomings of the PY solution. The form factors $F_i(k)$
were calculated with a model of spherical three-layer particles, and a
difference between HS and scattering core diameters, with $\sigma _i>\sigma
_i^{{\rm scatt}}$, was admitted.

Since our aim is not an exhaustive analysis of all data by Duits {\it et al.}
\cite{Duits91}, in Figure 2 we consider only two representative cases, with $%
\eta =0.28$ and $t=51.6$ $^{\circ }$C and $35.4$ $^{\circ }$C, respectively.
The $S_{{\rm M}}(k)$ of the first sample (Figure 2a) is typically
``HS-like'', whereas at low $k$ the curve of Figure 2b has an upswept shape
characteristic of strong attraction. For simplicity, we calculate the form
factors with a model of homogeneous spherical scattering cores. We have
checked that this choice is essentially equivalent to the three-layer one,
if appropriate average contrast and diameter ( $\left\langle \sigma ^{{\rm %
scatt}}\right\rangle =39.8$ nm ) are used, as suggested by Ref. \cite
{Duits91}.

In Figure 2a we first fit the experimental data by employing the
monodisperse SHS-MSA model (all particles having the same size and
stickiness; $s_\sigma =0$), with two free parameters, $T^{*}$ and $\sigma
>\left\langle \sigma ^{{\rm scatt}}\right\rangle .$ The best fit results
are: $(T^{*},\sigma )=(4.2,\ 45.6),$ with the value of $T^{*}$ confirming
that the system is in the ``repulsive'' regime. Clearly, this model
overestimates the first peak height and does not reproduce the rising part
of the curve adequately. Both these drawbacks can be overcome by taking size
polydispersity into account. Indeed, both Model I and II yield excellent
agreement, and their results are nearly identical, indicating that the
choice of stickiness distribution is irrelevant in this HS-like case. To use
only two free parameters, i.e. $\left\langle \sigma \right\rangle
>\left\langle \sigma ^{{\rm scatt}}\right\rangle $ and $s_\sigma $, for
Model I and II we fix $T^{*}$ on the value found with the monodisperse
model. The best fit results are: $(\left\langle \sigma \right\rangle
,s_\sigma )=(44.9,\ 0.12)$ for Model I, and $\left( 44.2,\ 0.15\right) $ for
Model II (using three free parameters, i.e. $T^{*},\left\langle \sigma
\right\rangle ,s_\sigma $, does not alter our qualitative results). We
remark that both our estimates of $s_\sigma $ are consistent with the value $%
0.11$ of Ref. \cite{Duits91}. Moreover, comparison with Figure 6 of Ref. 
\cite{Duits91} shows that Model I improves over its PY counterpart of Duits 
{\it et al.} This may be due to our use of a much larger number $p$ of
components to represent the size distribution.

In Figure 2b we use a similar strategy. Here, however, we have found
impossible to satisfactorily fit the entire experimental curve with the
monodisperse model. Hence, we have forced $T^{*}$ to the value $2.3$ which
best reproduces the data within the restricted $k$-range $(0.03,\ 0.15)$ nm$%
^{-1}$ (note that this value of $T^{*}$ indicates that the system is in the
``attractive'' regime). Then, upon fitting the complete set of data with one
free parameter, we find $\left\langle \sigma \right\rangle =\ 43.0$ nm for
the monodisperse model. On fixing $T^{*}=2.3$, the best results for Model I
and II are: $(\left\langle \sigma \right\rangle ,s_\sigma )=(41.9,\ 0.12)$
and $\left( 41.5,\ 0.13\right) $, respectively. Once again we note the
importance of the polydispersity effects, which lower the first peak height
correctly and improve the general agreement. Both Models are satisfactory,
although Model II appears to be slightly superior, indicating, perhaps, that
the difference of stickiness distribution is already becoming significant.
However, this conclusion cannot be definitive. In fact, further calculations
(not reported here) under different conditions, yield ambiguous results.
Moreover, Figure 1 shows that the differences between Model I and II are
small at low $s_\sigma $ values, as unfortunately occurs in the colloidal
system investigated by Duits {\it et al.} \cite{Duits91}. On the other hand,
our results suggest the opportunity of further experimental investigations
on similar fluids with larger polydispersity, where an analysis of $S_{{\rm M%
}}(k)$ at low $k$ could discriminate between different SHS Models and
provide useful information about the unknown real stickiness distribution.

\section{Conclusions}

The present study on a new polydisperse SHS model has once again shown the
importance of taking polydispersity effects into account when experimental
scattering data are analyzed. We have found that even a small degree of size
polydispersity, $s_\sigma \simeq 0.1$, can strongly modify $S_{{\rm M}}(k)$
with respect to the monodisperse case. We have pointed out the existence of
``repulsive'' and ``attractive'' regimes for $S_{{\rm M}}(k),$ with a
crossover interpreted in terms of a ``generalized Boyle temperature'',
dependent on the degree of polydispersity. In the presence of
polydispersity, a crossover between these regimes can induced not only by a
change of $T^{*}$, but also by a change of $s_\sigma $ (at fixed $T^{*}$).

In addition to size polydispersity, we have considered two simple stickiness
distributions (Model I and II). Our results indicate that the particular
choice of stickiness distribution is irrelevant in fluids with small size
polydispersity and weak coupling (i.e. weak attraction and high
temperature). On the contrary, stickiness polydispersity effects can be
revealed from the low-$k$ behaviour of $S_{{\rm M}}(k)$ when size
polydispersity is large and coupling is strong (strong adhesive forces and
low temperature).

As an application, we have employed Model I and II to analyze experimental
scattering data for a colloidal fluid with short-range attractions. Our best
fit results are satisfactory and suggest the importance of the stickiness
polydispersity (Model II). However, the considered experimental system lies
in a regime where it is impossible to unambigously discriminate between our
two Models. New scattering studies on fluids with larger size polydispersity
and stronger adhesive forces would be desirable to get more insight into
this problem.

{\bf Acknowledgements}

Partial financial support by the Italian INFM (Istituto Nazionale di Fisica
della Materia) is gratefully acknowledged.

\begin{figure}[tbp]
\centerline{ \epsfxsize=3.5truein \epsfysize=3.5truein
\epsffile{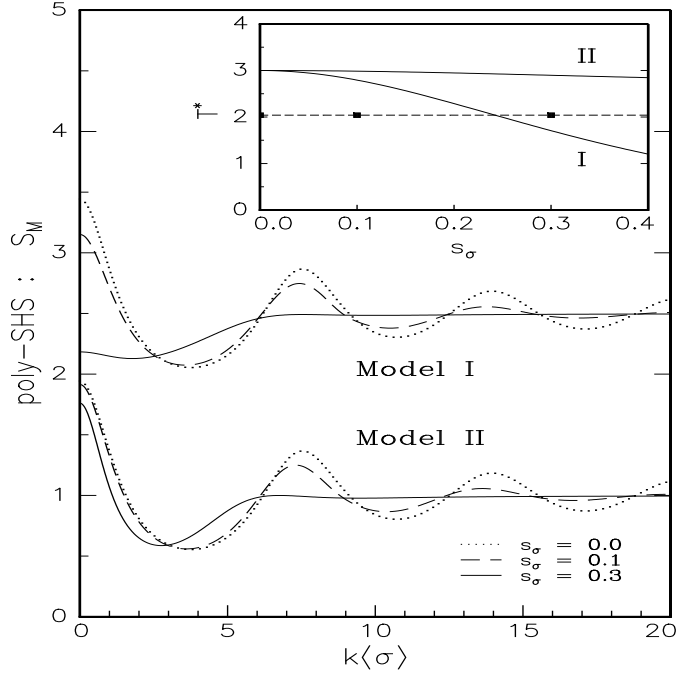} }
\caption{ Structure factor $S_{{\rm M}}(k)$ of the SHS-MSA Models I and II at $%
T^{*}=2.04$ and $\eta =0.2$, for several degrees of polydispersity $s_\sigma$ 
(the curves of Model I are shifted upwards by 1.5 units to avoid overlapping).
In the inset, the solid curves represent $T_{{\rm B,F}}^{*}(s_\sigma )$
of the two Models, the dashed line corresponds to the isotherm $T^{*}=2.04$
and the black squares on it are the considered states. }
\end{figure}

\begin{figure}[tbp]
\centerline{ \epsfxsize=3.5truein \epsfysize=3.5truein
\epsffile{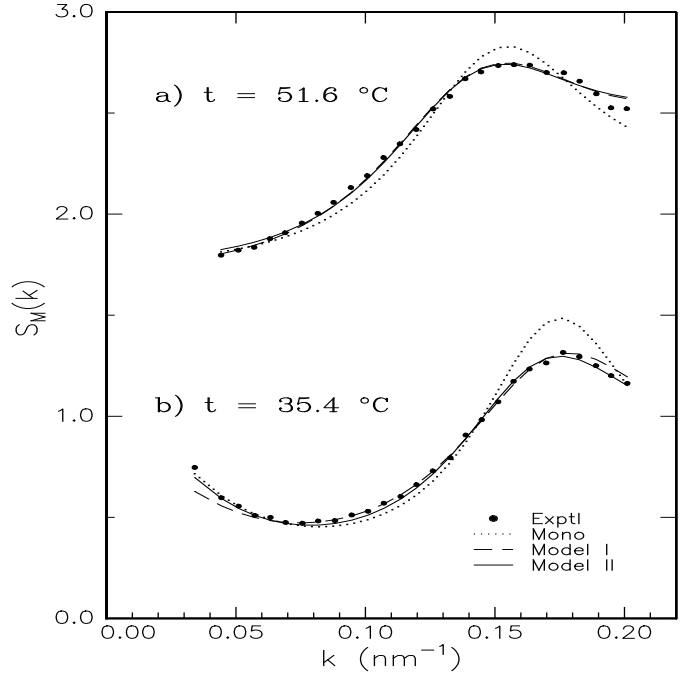} }
\caption{ Fit of experimental neutron scattering data for $S_{{\rm M}}(k)$ at $\eta
=0.28$ and two different temperatures (taken from Ref. [3]),
 by using monodisperse and polydisperse SHS-MSA models (the curves of
 part (a) are shifted upwards by 2.5 units). }
\end{figure}

\end{document}